\documentclass[pra,twocolumn,superscriptaddress,10pt,noshowpacs]{revtex4}
\usepackage[portuguese]{babel}
\usepackage[T1]{fontenc}
\usepackage[utf8]{inputenc}
\usepackage{graphicx,epstopdf}
\usepackage{amsmath}

\usepackage[pdftex,plainpages=false,colorlinks=true,citecolor=blue,linkcolor=blue,urlcolor=blue,filecolor=green,bookmarksopen=true]{hyperref}
\usepackage{amsfonts}
\usepackage{bbm}
\usepackage{amssymb}
\usepackage{color}
\usepackage{latexsym}
\usepackage{caption}
\usepackage{subcaption}
\usepackage{times,txfonts}

\begin{document}

\title{Effects of Lorentz violation in superconductivity}

\author{J. Furtado}
\affiliation{Centro de Ci\^{e}ncias e Tecnologia, Universidade Federal do Cariri, 63048-080, Juazeiro do Norte, Cear\'{a}, Brazil}
\email{job.furtado@ufca.edu.br}

\author{R. M. M. Costa Filho}
\affiliation{Universidade Federal de Alagoas - UFAL, 22290-180, Macei\'{o} - AL, Brasil}

\author{A. F. Morais}
\affiliation{Departamento de F\'{i}sica, Universidade Regional do Cariri, 57072-270, Juazeiro do Norte, Cear\'{a}, Brazil}

\author{I. C. Jardim}
\affiliation{Departamento de F\'{i}sica, Universidade Regional do Cariri, 57072-270, Juazeiro do Norte, Cear\'{a}, Brazil}

\date{\today}

\begin{abstract}
This paper presents the effects of Lorentz violation in superconductivity. Constructing a Lorentz-Violating Ginzburg-Landau theory of superconductivity we discuss the influence of the Lorentz-Violating tensors $\hat{k}_{c}^{ij}$ and $\hat{k}_a^i$ in the London's depth penetration, in the coherence length and critical magnetic field.  
\end{abstract}

\maketitle

\section{Introduction}

In the last years, possible extensions of the Standard Model (SM) have been studied and, in this context, Lorentz and CPT symmetries breaking are now considered as an important topic of discussion \cite{Kos1, Kos2, Kos3, Kos4, Colladay:1996iz, Colladay:1998fq}. Usually, the Lorentz symmetry is broken by introducing privileged directions in space-time, expressed through some additive terms which are proportional to small constant vectors or tensors. The most well-established model that considers the effects of Lorentz and CPT symmetry violation is the Standard Model Extension (SME) \cite{Colladay:1996iz, Colladay:1998fq}, an effective field theory that includes in its lagrangian all possible terms that violate Lorentz and CPT symmetries.

An extension of the scalar sector considering Lorentz-Violating effects was proposed by Kostelecky \cite{Edwards:2018lsn}. Such a model present a general effective scalar field theory in any spacetime dimension containing explicit perturbative spin-independent Lorentz violating operators of arbitrary mass dimension. The importance of this construction relies on the fact that the great majority of the fundamental particles of the SM have spin, being the Higgs boson the only example of a fundamental spinless particle in the SM. In spite of the minor role played by the scalar sector of QED (sQED), in comparison to the strong interaction, in describing the coupling between mesons, it was argued \cite{Edwards:2019lfb} that a Lorentz-violating extension of sQED could be an effective way of treating small CPT deviations in neutral-mesons oscillations.

Potential applications of Lorentz and CPT violating models range from quantum field theory to condensed matter physics. Such interface between Lorentz-violating models and condensed matter physics have received great attention specially in the context of Weyl semimetals \cite{Assuncao:2015lfa, Gos1, Gos2, Gos3}, superconductivity \cite{Bazeia:2016pra}, graphene with anisotropic scaling \cite{Katsnelson:2012cz}, dark matter and black holes analog models \cite{Baym:2020uos, Pereira:2009vb}, among others.

The Ginzburg-Landau (GL) theory of superconductivity is an effective quantum field theory construction that can explain several important aspects of superconductivity \cite{Ginzburg:1950sr}. Although initially thought as a phenomenological theory, the GL theory can be interpreted as a limiting case of the Bardeen-Cooper-Schrieffer (BCS) theory of superconductivity \cite{Bardeen:1957mv}. Superconductivity in Weyl semimetals \cite{Zhou:2015qka, Bednik:2015tha, Wei:2014vsa} as well as in carbon-based nanostructures such as graphene and fullerene \cite{Roy:2013aya, Cohnitz:2017vsr} are topics often addressed in the literature due to their possible applications in technological improvement. However the study of superconductivity in a more theoretical framework such as superconductivity in astrophysics \cite{Madsen:1999ci, Bonanno:2011ch}, cosmology \cite{Ebert:2007ey, Gao:2012aw} and in high energy physics \cite{Herzog:2009xv, Ghoroku:2019trx}, for example, has been intensively discussed in the last years.

In this study we consider the Lorentz-violating complex scalar sector proposed by Kostelecky \cite{Edwards:2018lsn} coupled minimally with the gauge field, the usual Maxwell term as the gauge sector and $\lambda|\phi|^4$ promoting the spontaneous symmetry breaking. The Lorentz-violating parameters can be interpreted as defects or layers in the superconductor, giving rise to an anisotropy in the system.   

This paper is organized as follows: in the next section we present our model, discussing the discrete symmetries and the Lorentz-violating extension of the Ginzburg-Landau theory for superconductivity. In the section III we discuss the superconducting case for the contributions from $\hat{k}_c^{\mu\nu}$ and $\hat{k}_a^{\mu}$. Finally, in section IV we highlight our conclusions.

\section{Model}
The model we are considering consists of the Lorentz-violating complex scalar sector proposed by Kostelecky \cite{Edwards:2018lsn} coupled minimally with the gauge field, the usual Maxwell term as the gauge sector and $\lambda|\phi|^4$ potential promoting the spontaneous symmetry breaking. Hence the Lagrangian describing the system is 

\begin{eqnarray}\label{lagrangian1}
\nonumber\mathcal{L}&=&G^{\mu\nu}(D_{\mu}\phi)^*D_{\nu}\phi-m^2\phi^*\phi-\lambda(\phi^*\phi)^2\\
&&-\frac{i}{2}[\phi^*\hat{k}_a^{\mu}D_{\mu}\phi-\phi\hat{k}_a^{\mu}(D_{\mu}\phi)^*]-\frac{1}{4}F^{\mu\nu}F_{\mu\nu},
\end{eqnarray}
where $D_{\mu}=\partial_{\mu}-ieA_{\mu}$ is the usual covariant derivative, $F_{\mu\nu}=\partial_{\mu}A_{\nu}-\partial_{\nu}A_{\mu}$ and the tensor $G^{\mu\nu}=g^{\mu\nu}+(\hat{k}_c)^{\mu\nu}$ is composed by the Minkowski metric tensor $g^{\mu\nu}=diag(1,-1,-1,-1)$ and a Lorentz-violating constant tensor $(\hat{k}_c)^{\mu\nu}$. The tensors $(\hat{k}_c)^{\mu\nu}$ and $\hat{k}_a^{\mu}$ promotes the violation of the Lorentz invariance by breaking the equivalence between particle and observer transformations. Such tensors, assumed to be constant, imply the independence of the space-time position, which yields translational invariance assuring the conservation of momentum and energy. Note that while the tensor $(\hat{k}_c)^{\mu\nu}$ is dimensionless, the tensor $\hat{k}_a^{\mu}$ has dimension of mass. 

Regarding the analysis of the discrete symmetries on the LV tensors $\hat{k}_a^{\mu}$ and $(\hat{k}_c)^{\mu\nu}$, the results are summarized in the table. As we can see the tensor $\hat{k}_a^{\mu}$ is CPT-odd while $(\hat{k}_c)^{\mu\nu}$ is CPT-even. The PT symmetry is always preserved, however, effects of CP violation can be saw with the $\hat{k}_a^{0}$ and $(\hat{k}_c)^{0i}$ components. It is important to highlight here that the $\hat{k}_a^i$ violates charge, partity and time reversal symmetries simultaneously.

\begin{table}[h!]
\begin{tabular}{|l|l|l|l|l|}
\hline
 & C & P & T & CPT \\ \hline
\,\,\,\,\,\,\,\,\,\,\,\,$\hat{k}_a^{0}$ & - & +  & +  & \,\,\,\,- \\ \hline
\,\,\,\,\,\,\,\,\,\,\,\,$\hat{k}_a^{i}$ & - & -  & -  & \,\,\,\,- \\ \hline
$(\hat{k}_c)^{00}$, $(\hat{k}_c)^{ij}$ & + & + & + & \,\,\,+ \\ \hline
\,\,\,\,\,\,\,\,\,\,$(\hat{k}_c)^{0i}$ & + & - & - & \,\,\,+ \\ \hline
\end{tabular}
\end{table}

In the static condition the Lagrangian becomes:
\begin{eqnarray}\label{lagrangian2}
    \nonumber\mathcal{L}&=&G^{ij}(D_i\phi)^*(D_j\phi)-\mu^2\phi^*\phi-\lambda(\phi^*\phi)^2\\
    &&-\frac{i}{2}[\phi^*\hat{k}_a^{i}D_{i}\phi-\phi\hat{k}_a^{i}(D_{i}\phi)^*]-\frac{1}{4}F_{ij}F^{ij}.
\end{eqnarray}
Hence, $-\mathcal{L}$ is the Lorentz-Violating Ginzburg-Landau free energy. Note that the mass parameter $m^2$ was written as $\mu^2$ which is now considered as a temperature-dependent system parameter $\mu^2=a(T-T_c)$ near the critical temperature $T=T_c$. In this context, $\phi$ is the macroscopic many-particle wave function. The physical interpretation of $\phi$ as a many-particle wave function is justified by the Bardeen-Cooper-Schrieffer (BCS) theory, according to which, under certain conditions, there is an attractive force between electrons, and field quanta are electron pairs, which are, of course, bosons. At low temperatures, the field quanta fall into the same quantum state (Bose-Einstein Condensation \cite{Furtado:2020olp, Casana:2011bv}), and because of this, a many-particle wave function $\phi$ may be used to describe the macroscopic system.

The equations of motion for the complex scalar field and gauge field are:
\begin{eqnarray}
    \nonumber&&G^{ij}\left[-\partial_i\partial_j-2ie A_j\partial_i+e^2A_iA_j-ie(\partial_jA_i)\right]\phi^*-\mu^2\phi^*\\
    \nonumber&&-2\lambda|\phi|^2\phi^*-e\hat{k}_a^iA_i\phi^*+i\hat{k}_a^i\partial_i\phi^*+\frac{i}{2}(\partial_{i}\hat{\kappa}_{a}^{i})\phi^{*}\\ &&-\partial_{i}(\hat{k}_c)^{ij}\left(\partial_{j}+ieA_{j}\right)\phi^{*}=0\\
    \nonumber&&ieG^{mi}(\phi^*\partial_i\phi-\phi\partial_i\phi^*)+2e^2G^{mi}A_i|\phi|^2-2e\hat{k}^{m}_{a}|\phi|^{2}\\
    &&+\Box A^m-\partial_k\partial^mA^k=0.
\end{eqnarray}

The ground state is obtained by minimizing the potential $V(\phi^*,\phi)$, defined as, 
\begin{equation}
    V(\phi^*,\phi)=\mu^2|\phi|^2+\lambda|\phi|^4.
\end{equation}
At $T>T_c$, $\mu^2>0$ and the minimum free energy is at $|\phi|=0$. But when $T<T_c$, $\mu^2<0$ and the minimum free energy is at
\begin{equation}
    |\phi|^2=-\frac{\mu^2}{2\lambda}>0.
\end{equation}
In all cases considered above $\lambda >0$. 

Note that the lagrangian (\ref{lagrangian2}) possess an obvious $U(1)$ symmetry, so that
\begin{eqnarray}
    \phi&\rightarrow& \phi'=e^{-i\alpha}\phi\\
    A^{i}&\rightarrow& A'^{i}=A^{i}-\frac{1}{e}\partial^{i}\alpha,
\end{eqnarray}
with $\alpha\in\mathbb{R}$. Noether's theorem states that for any given continuous symmetry there is a conserved quantity in connection. The probability current yields
\begin{equation}
    j^{i}(x)=-iG^{ij}(\phi^*\partial_{j}\phi-\phi\partial_{j}\phi^*)+[-2eG^{ij}A_j+(\hat{k}_a)^i]|\phi|^2.
\end{equation}
When $T<T_c$ ($\mu^2<0$) and $\phi$ varies only very slightly over the sample, the second term dominates over the first, so that,
\begin{eqnarray}
\vec{j}=-\Gamma^2\vec{G}+\frac{\Gamma^2}{2e}\vec{\kappa}.
\end{eqnarray}
with $\vec{G}=G^{ij}A_j$, $\vec{\kappa}=(\hat{k}_a)^i$ and
\begin{eqnarray}
\Gamma^2=-\frac{e\mu^2}{\lambda}.
\end{eqnarray}
The constant $\Gamma$ is positive definite. From now on we will be working only with the case when $T<T_c$, therefore, no phase phase transitions are considered. The above equation is the so called London equation modified by the presence of the Lorentz-violating tensor $\hat{k}_c^{ij}$ and vector $\hat{k}_a^i$. The electric field is defined as $\vec{E}=-\partial\vec{A}/\partial t=0$ and the Ohm's law defines resistance by $\vec{E}=R\vec{j}$, so, we have two possible solutions, i.e., $R=0$ or $\vec{j}=0$. 

A non-trivial solution for the insulating case ($\vec{j}=0$) occurs when $\vec{\kappa}\neq \vec{0}$. Thus, without loss of generality, let us consider $\hat{k}_c^{ij}=0$, so that, $\vec{G}=\vec{A}$. Then, the insulation condition $\vec{j}=\vec{0}$ gives us

\begin{equation}\label{NC1}
    \vec{A}=\frac{1}{2e}\vec{\kappa}.
\end{equation}
This condition imposes that the potential vector and the Lorentz-violating vector $\vec{\kappa}$ must be collinear. The static condition imposes a zero electric field. Taking curl in both sides of (\ref{NC1}) we have
\begin{equation}\label{NC2}
    \vec{B}=\frac{1}{2e}(\nabla\times\vec{\kappa}).
\end{equation}
As an immediate consequence note that if $\vec{\kappa}$ is a constant or irrotational vector then (\ref{NC2}) renders $\vec{B}=0$. From Faraday's law, we confirm that the magnetic field must be static while the Amp\`{e}re-Maxwell equation leads to $\nabla\times\vec{B}=0$. Now, taking curl in (\ref{NC2}) we obtain 
\begin{equation}
    \nabla(\nabla\cdot\vec{\kappa})-\nabla^2\vec{\kappa}=0.
\end{equation}
The above equation sets the condition for $\vec{\kappa}$ in order to guarantee an absence of electric current inside the material.

\section{Superconducting case: $R=0$}

In this section we will study the superconducting case ($R=0$) for both $\hat{k}_c^{ij}$ and $\hat{k}_a^i$ Lorentz-violating terms. In the context of GL theory of superconductivity, three important properties can be obtained, namely, the coherence length, the critical magnetic field, and London's penetration depth. 

In order to obtain the coherence length of the superconductor we have to solve the equation of motion for $\phi$ considering the following boundary conditions: in the surface of the superconductor we must have $\phi=0$ and deep inside the material $\phi=\phi_{max}$, so that $\phi_{max}$ is practically constant, which means that $\partial_i\phi_{max}=0$. Hence, the highest density of superconducting electrons occurs as usual at $|\phi_{max}|^2=-2\mu^2/\lambda$. Thence, it is straightforward to find that
\begin{equation}
    |\phi|^2=|\phi_{max}|^2\tanh^2\left({\frac{\sqrt{2}x}{2\xi}}\right),
\end{equation}
with the coherence length $\xi$ being given by
\begin{equation}
    \xi=\sqrt{\frac{\hbar^2}{2m^*|\mu^2|}},
\end{equation}
where $m^*=2m_e$, being $m_e$ the effective electron's mass. The factor of 2 takes into account the fact that the $\phi$ field stands for Cooper pairs. 

%\begin{figure}[h!]
%\begin{center}
%\includegraphics[scale=0.4]{SCfig2.eps}
%\caption{Coherence length for the LV-superconductivity}
%\label{fig2}
%\end{center}
%\end{figure}

The critical magnetic field, i.e., the maximum value of $H$ that does not destroy the superconducting state, is not modified by the presence of the LV contributions. So that the behaviour of the magnetic field as a function of temperature does not change (type-I superconductor), and the value of the critical magnetic field is given by:
\begin{equation}
    H_c=\sqrt{\frac{2\mu^4}{\mu_0\lambda}}.
\end{equation}
Note that since $\lambda>0$ and $\mu^4>0$ it follows that $H_c>0$.

Differently from the coherence length and the critical magnetic field, London's penetration depth is modified by the presence of the LV contributions $\hat{k}_c^{ij}$ and $\hat{k}_a^i$. For the sake of clarity, we will consider the contributions from $\hat{k}_c^{ij}$ and $\hat{k}_a^i$ individually.

\subsection{Contribution from $\hat{k}_c^{ij}$}

Let us consider the case when $\hat{k}_a^{i}=0$, which means that we are studying only the effects of $k_c^{ij}$. When $R=0$ and $\hat{k}_c^{ij}=Kg^{ij}$ with $K$ a constant, the Amp\`{e}re-Maxwell equation gives us $\nabla\times\vec{B}=\vec{j}$, from which we take a curl remembering that $\nabla\cdot\vec{B}=0$, which yields,
\begin{equation}
    -\nabla^2\vec{B}=-\Gamma^2(1+K)\vec{B}.
\end{equation}
If $K=0$, hence,
\begin{equation}\label{e1}
    \nabla^2\vec{B}=\Gamma^2\vec{B}.
\end{equation}
In this case, the solution for $B^i=B_0e^{-\Gamma x^i}$, specifies the usual London's depth penetration, which is proportional to $\Gamma^{-1}$. The London's depth penetration can also be interpreted as a consequence of a massive photon. 

\begin{figure}[h!]
\begin{center}
\includegraphics[scale=0.4]{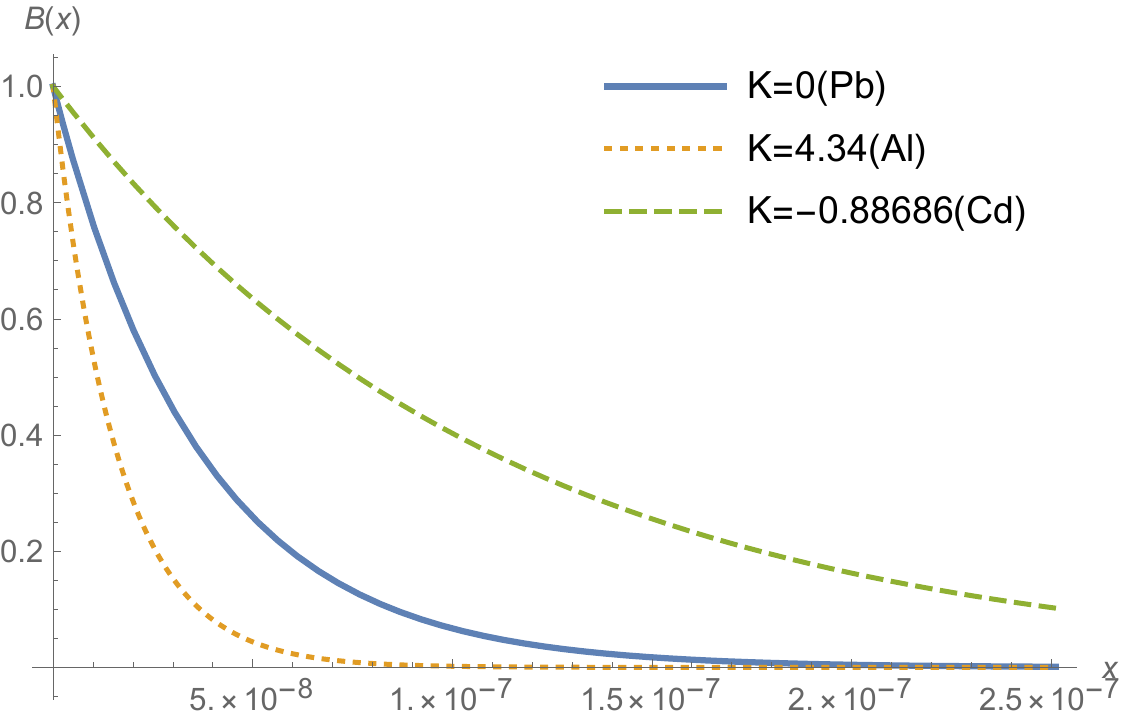}
\caption{Effect of a constant $\hat{k}_c^{ij}=Kg^{ij}$ on the behaviour of the magnetic field inside the superconductor. We set $B_0=1T$ and $\Gamma^{-1}=37nm$. For this plot we have considered three values for $K$, namely, $K=0$, $K=4.34$ and $K=-0.88$.}
\label{fig1}
\end{center}
\end{figure}

Considering $K$ constant but non-vanishing the magnetic field behaves as in fig (\ref{fig1}). Note that we can interpret the constant $K$ as a parameter that can define different isotropic superconductors depending on the value of $K$. In fig (\ref{fig1}) we have considered $\Gamma^{-1}=37nm$, which correspond to the London's penetration depth for a $Pb$ superconductor, and $B_0=1T$. As we can see, considering $K=4.34$ we can describe a pure $Al$ superconductor and if we set $K=-0.88$ we obtain the London's penetration depth for a pure $Cd$ superconductor. Also, as the value of $K$ decreases, the magnetic field penetrates deeper into the superconductor.

However, if we consider a position-dependent $\hat{k}_c^{ij}=\hat{k}_c^{ij}(x^m)$ the equation describing the behaviour of the magnetic field becomes:
\begin{eqnarray}
\nonumber\nabla^2\vec{B}&=&\Gamma^2\vec{B}+\Gamma^2\left[(\partial_2\bar{A}_3-\partial_3\bar{A}_2)\hat{x}\right.\\
&&+\left.(\partial_3\bar{A}_1-\partial_1\bar{A}_3)\hat{y}+(\partial_1\bar{A}_2-\partial_2\bar{A}_1)\hat{z}\right],
\end{eqnarray}
where $\bar{A}^i=\hat{k}_c^{ij}(x,y,z)A_j$. But since $\vec{B}=\nabla\times\vec{A}$, we can also write an equation for the potential vector inside the superconductor as 
\begin{equation}
    \nabla^2\vec{A}=\Gamma^2\left(\vec{A}+\stackrel{\leftrightarrow}{k}_c\cdot\vec{A}\right).
\end{equation}
For this case, for the sake of simplicity, we will consider two configurations of $\hat{k}_c^{ij}$: a diagonal configuration and an off-diagonal one. For the diagonal configuration, we have
\begin{eqnarray}
\hat{k}_c^{ij}(x)=\left(\begin{array}{ccc}
    K^{11}(x) & 0 & 0  \\
    0 & K^{22}(x) & 0\\
    0 & 0 & -K^{11}(x)-K^{22}(x)
\end{array}\right),
\end{eqnarray}
where we are taking into account the traceless property and the functions $K^{11}(x)$ and $K^{22}(x)$ controls the anisotropy of the system.
The magnetic field inside the superconductor is described by three coupled differential equations
\begin{subequations}\label{kcdiag}
\begin{eqnarray}
\partial_3^2B_z&=&\Gamma^2B_z+\Gamma^2\left\{\partial_1[K^{22}(x)A_2]-\partial_2[K^{11}(x)A_1]\right\}\\
\partial_2^2B_y&=&\Gamma^2B_y+\Gamma^2\left\{\partial_3[K^{11}(x)A_1]-\partial_1[K^{33}(x)A_3]\right\}\\
\partial_1^2B_x&=&\Gamma^2B_x+\Gamma^2\left\{\partial_2[K^{33}(x)A_3]-\partial_3[K^{22}(x)A_2]\right\}.
\end{eqnarray}
\end{subequations}

The off-diagonal configuration, in which $\hat{k}_c^{ij}$ is given by
\begin{eqnarray}
\hat{k}_c^{ij}(x)=\left(\begin{array}{ccc}
    0 & K^{12}(x) & K^{13}(x)  \\
    K^{12}(x) & 0 & K^{23}(x)\\
    K^{13}(x) & K^{23}(x) & 0
\end{array}\right),
\end{eqnarray}
exhibits the symmetry property of $\hat{k}_c^{ij}$ and the functions $K^{12}(x)$, $K^{13}(x)$ and $K^{23}(x)$ controls the anisotropy of the system. Similarly to the diagonal configuration, the magnetic field inside the superconductor is described by three coupled differential equations 
\begin{subequations}\label{kcoffdiag}
\begin{eqnarray}
\nonumber\partial_3^2B_z&=&\Gamma^2B_z+\Gamma^2\left\{\partial_1[K^{12}(x)A_1+K^{23}(x)A_3]\right.\\
&&\left.-\partial_2[K^{12}(x)A_2+K^{13}(x)A_3]\right\}\\
\nonumber\partial_2^2B_y&=&\Gamma^2B_y+\Gamma^2\left\{\partial_3[K^{12}(x)A_2+K^{13}(x)A_3]\right.\\
&&\left.-\partial_1[K^{13}(x)A_1+K^{23}(x)A_2]\right\}\\
\nonumber\partial_1^2B_x&=&\Gamma^2B_x+\Gamma^2\left\{\partial_2[K^{13}(x)A_1+K^{23}(x)A_2]\right.\\
&&\left.-\partial_3[K^{12}(x)A_1+K^{23}(x)A_3]\right\}.
\end{eqnarray}
\end{subequations}

These equations for the magnetic field ((\ref{kcdiag}) and \ref{kcoffdiag}) can be used to be modeling a great variety of anisotropic superconductors. The free parameters, i.e., the independent components of $\hat{k}_c^{ij}$ can be interpreted basically in two different ways. In the first interpretation, the free parameters are related to the properties of the lattice itself while in the second interpretation, the independent components of $\hat{k}_c^{ij}$ are related to the process of producing the superconductor, for instance, a multiple layer superconductor. 

An interesting third interpretation can be given to the free parameters. It is also possible to be modeling topological defects in the superconductor with the free parameters set as delta or sharp Gaussian functions.

\subsection{Contribution from $\hat{k}_a^{i}$}

Let us consider the case when $\hat{k}_c^{ij}=0$. In this case $G^{ij}=g^{ij}$, which means that we are studying the effects of $k_a^{i}$. When $R=0$, the Amp\`{e}re-Maxwell equation gives us $\nabla\times\vec{B}=\vec{j}$, from which we take a curl remembering that $\nabla\cdot\vec{B}=0$, which yields,
\begin{equation}\label{Londonequationk_a}
    -\nabla^2\vec{B}=-\Gamma^2\vec{B}+\frac{\Gamma^2}{2e}(\nabla\times\vec{\kappa}).
\end{equation}
If $\vec{\kappa}$ is a constant or irrotational vector field, then $\nabla\times\vec{\kappa}=0$, hence we obtain (\ref{e1}). 

Similarly to the case studied previously, we can also consider a position-dependent $\vec{\kappa}=\vec{\kappa}(\vec{x})$ such as $\nabla\times\vec{\kappa}\neq0$. In this case we are discussing different anisotropic superconductors other than those discussed previously. The main difference is that the free parameters controlling the anisotropy, i.e., the components of $\vec{\kappa}$, do not couple directly to the potential vector, meaning that they act as an external source for the magnetic field.

\section{Final Remarks}

In this paper, we presented the effects of Lorentz violation in superconductivity. Constructing a Lorentz-Violating Ginzburg-Landau theory of superconductivity we discuss the influence of the Lorentz-Violating tensors $\hat{k}_{c}^{ij}$ and $\hat{k}_a^i$ in the London's depth penetration, in the coherence length and in the critical magnetic field. 

The model is inspired by the Lorentz-violating extension of the complex scalar sector of the Standard Model of particle physics. The constructed Lorentz-Violating Ginzburg-Landau model can be used to describe a great variety of anisotropic superconductors. The anisotropy is controlled by the tensor $\hat{k}_c^{ij}$ and the vector $\vec{\kappa}$. 

We have shown that the presence of the Lorentz-violating contributions does not affect the coherence length and the critical magnetic field that sets the superconducting state. However, the London's depth penetration is modified by the presence of the Lorentz-violating parameters. The presence of the Lorentz-violating terms strongly influences the behaviour of the magnetic field inside the superconductor.

The tensor $\hat{k}_c^{ij}$ couples directly with the potential vector and, in the case of being position-dependent, it can be interpreted as a consequence of the properties of the lattice of the material as well as a property emergent from the process of constructing the superconductor, as a multilayered superconductor. On the other hand, the vector $\vec{\kappa}$ does not couple to the potential vector. In order to give rise to an anisotropy, the vector $\vec{\kappa}$ must be irrotational. Besides, we have shown that the vector $\vec{\kappa}$ plays the role of an external source for the magnetic field inside the superconductor. 

Some future perspectives for this work may include the study of the Abrikosov vortex in the present context and the consideration of holes inside the superconductor. We can also consider non-minimal couplings between the complex scalar field and the gauge field.

\end{document}